\documentclass[letterpaper,twocolumn,10pt]{article}

\usepackage{usenix-2020-09}

\usepackage{placeins}
\usepackage{papermacros}
\usepackage{hyperref}
\usepackage{multirow}

\usepackage{array}
\usepackage{caption}
\newcommand{\head}[2]{\multicolumn{1}{>{\centering\arraybackslash}p{#1}}{#2}}
\newcommand{\projname}{SpraySense\xspace}

\microtypecontext{spacing=nonfrench}
\title{{\projname}: Sensing Gray Failures through Packet Spraying Discrepancies}

\renewcommand{\paragraph}[1]{\textbf{#1. }}

\newcommand{\simulation}[0]{\xspace\lbrack{}Simulation\rbrack\xspace}
\newcommand{\testbed}[0]{\xspace\lbrack{}Testbed\rbrack\xspace}
\author{
Jakob Krebs\\{\rm Technion}
\and
Daniel Amir\\{\rm Technion}
\and
Shir Landau Feibish\\{\rm University of Haifa}
\and
Mark Silberstein\\{\rm Technion/NVIDIA}
}
\begin{document}
\maketitle\sloppy

\begin{abstract}
Distributed machine learning (ML) training has become a dominant workload in modern data center networks, operating at massive scale with clusters comprising tens to hundreds of thousands of GPUs. The scale of these networks makes failures, and particularly gray failures, inevitable. Gray failures can significantly degrade both network and application performance, yet they are notoriously difficult to detect, localize, and debug. 
To meet the performance demands of ML workloads, adaptive routing is widely deployed to maximize network utilization by dynamically spreading traffic across many paths. While adaptive routing increases network utilization, it also greatly intensifies the effect of gray failures. 
Prior work has either dismissed gray failures as negligible or proposed detection mechanisms that fail to scale, rendering these approaches increasingly impractical for large-scale clusters.

We present \textbf{SpraySense}, a passive gray failure detection system that leverages the statistical properties of adaptive routing and network load balancing. 
By combining these properties with flow-level information, SpraySense can identify failures before they significantly impact application performance, enabling preemptive rerouting and improving overall performance.
Importantly, this is achieved through passive observation of traffic spraying, without introducing additional load on the network. 
We evaluate SpraySense and show that it can detect and localize a single-link packet-drop-rate $1.5\%$ within a single iteration and as little as $0.5\%$ within 5 training iterations of Llama-3 70B in a 64 spine topology.

\end{abstract}

\section{Introduction}\label{sec:intro}

Distributed ML training has emerged as the dominant data center workload, justifying dedicated computing facilities.
Servers, the network, and even the buildings themselves are purpose-built for the specific demands of training frontier models.
Cluster networks now connect tens, or even hundreds of thousands of GPUs~\cite{colossus,meta-100k-collectives,meta-100k-training}, and comprise tens of thousands of switches and hundreds of thousands of links.
At this unprecedented scale, failures are inevitable.

Of particular concern are \emph{gray failures}, faults which evade traditional detection and localization at their source~\cite{grayfailure}.
Paths experiencing a gray failure silently drop a portion of traversing packets while appearing to the control plane to be in working order~\cite{netbouncer}.
Gray failures may arise from a variety of causes, such as contaminated fiber connectors causing elevated bit-error rates (BER), failing optical transceivers, or bit flips in switch packet buffers~\cite{zhuo2017understanding, fail-slow}.
In some cases, errors in switch logic only affect packets traversing specific paths, or only appear under particular load scenarios~\cite{pingmesh}. %
While rare, when these failures do occur, they can persist indefinitely.

The unique requirements of ML training have led to the broad deployment of specialized transport and load balancing techniques, such as in-network adaptive routing (AR)~\cite{adaptive-infiniband,dragonfly-plus}, to support the near-100\% utilization seen during network collectives. 
AR employs load-dependent packet spraying in uplink switches~\cite{dragonfly-plus, 2006-adaptive-routing, spectrumx-whitepaper, cisco-dlp},
distributing traffic evenly across the network fabric at the granularity of individual packets, and thus avoids ECMP collisions emerging due to low flow entropy of ML network workloads~\cite{drill, reps}.
As a result, AR has been broadly deployed in large-scale ML training clusters~\cite{spectrumx-whitepaper,spectrumx,colossus,supercharge-ai, cisco-dlp, cornelis}.

However, while AR ensures %
high network utilization, it amplifies the impact of gray failures: as packets are sprayed over all valid paths by the switch, any flow that has a valid path through a faulty link is likely to traverse it,  meaning that even a single failure can affect multiple flows.
Affected flows suffer from packet drops and subsequent time-consuming retransmissions, causing delays that rapidly propagate through the entire training cluster due to the bulk-synchronous nature of ML training~\cite{analysis-of-stragglers}.
Since gray failures stay undetected, every collective iteration will be affected.

\begin{figure}
    \centering
    \includegraphics[width=0.9\linewidth]{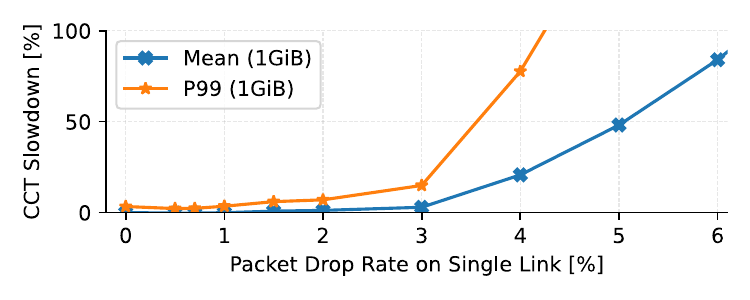}
    \vspace{-1em}
    \caption{\allreduce collective completion time (CCT) slowdown relative to failure-free network. 8 Spines; 1GiB collective size between 8 ranks.\simulation}
    \label{fig:allreduce-slowdown}
\end{figure}

To visualize the scale of the problem, \Cref{fig:allreduce-slowdown} shows the slowdown of an \allreduce collective due to packet drops on a leaf-spine link in a 2-level fat tree topology without redundant links.  \emph{A single link} with a gray failure inducing 3\% packet loss, i.e., only 0.375\% of packets in an affected sprayed flow, is sufficient to cause the 99th percentile collective completion time to be 14.7\% slower than without failures.

Traditional fault tolerance mechanisms can rapidly mitigate \emph{detected} failures~\cite{nvidia-convergence} by rerouting traffic, thus preventing further usage of affected paths. However, on their own, these mechanisms are ineffective against gray failures.
Since gray failures are invisible to the control plane, specialized mechanisms are required to detect and localize them as soon as they occur to allow the use of rapid-mitigation mechanisms.

\textbf{Existing approaches} to gray failure detection have fundamental weaknesses in the context of AR networks running ML training workloads. Path probing techniques~\cite{pingmesh, skeletonhunter, netbouncer} would be too costly and potentially ineffective. They detect gray failures by sending probing packets, but given low drop rates of gray failures, the volume of probes needed to detect them would add substantial network load. Moreover, probes might not even experience a failure that affects application packets, as they differ in size and structure, and traverse the fabric under different network conditions.
End-host monitoring~\cite{passive-failure-detection} requires controlling the packet path, which is not possible in switch-based AR. In-Network Telemetry techniques~\cite{p4-int} have scalability constraints as they require collecting detailed per-packet state from switches to a central location~\cite{pint, jia2020rapid, netseer, ufab}.

\textbf{Our approach}, which we call \emph{\projname}, is driven by the following observations: First, in a symmetric failure-free network, AR results in a symmetric spraying pattern.
In the absence of gray failures, this predictable pattern means that for each flow, the same number of packets traverse each possible path in expectation.
Second, ML training traffic is composed of large flows, ensuring that each flow's true spraying behavior closely approximates this expectation.
By counting how many packets of a given flow arrive from each spine switch, a destination leaf switch can determine whether the expected even distribution remains intact, or if some paths suffered packet drops due to a gray failure.
By comparing affected paths, failures can be rapidly localized and mitigated, even in cases with multiple parallel gray failures (\cref{sec:design:detection}).

This strategy addresses the weaknesses of existing approaches.
By leveraging intrinsic properties of AR, switches use \emph{the application traffic} directly to detect gray failures, without adding any load to the network.
By measuring within the network, path information is still available, allowing failures to be localized. And by using a simple signal, verified directly within switches, failures can be rapidly detected and mitigated without large centralized data collection overheads.

In practice, massive training networks are bound to contain preexisting failures and permanently disabled fabric links, which render them asymmetric~\cite{kokolis2025revisiting, alibaba-hpn}.
This asymmetry greatly complicates the predictability of spraying patterns, which now depend not only on the specific sources and destinations of each flow, but also on their \emph{relative timing}.
However, even under asymmetry, the spraying distribution of a single flow \emph{sent in isolation} is still easily predictable, given the network connectivity is already available in the switch's routing tables. 
We show that flow isolation, and predictable spraying behavior, can be recovered with minimal performance impact by using prioritization.
Each source leaf switch selects a single flow at a time for measurement, and prioritizes that flow's packets above all others.
This elevated prioritization is applied only during spraying, sufficient to ensure that the selected flow exhibits predictable spraying behavior (\cref{sec:asymmetry-predictability}).
At the same time, only a small portion of the total traffic originating from the leaf switch is prioritized, making performance impacts negligible (\cref{sec:evaluation:assumptions}).

\projname operates entirely within switches, combining line-rate data-plane measurements and lightweight control-plane logic. It adds no additional load to the network, requires no coordination between switches and only minimal application support (sending total flow sizes at the beginning of a collective), and enables precise failure localization.

We implement \projname in Tofino switches and evaluate it in our testbed with a full 2-level fat tree fabric using switch-based adaptive routing and ConnectX-6 DX NVIDIA NICs. We also implement it in an NS-3 packet-level simulation to demonstrate \projname's efficiency at larger scale. Our evaluation is based on real network collectives executed using the UCC collective library~\cite{ucc}.

For a network with 64 spines, \projname detects gray failures resulting in a 1.5\% loss rate with perfect accuracy within a single training iteration of Llama-3 70B~\cite{llama3-herd}, and a 0.5\% loss rate within 5 iterations (\cref{sec:eval:detection}).
The detection covers all paths used by the application, is robust to network noise caused by competing flows or congestion control, and can even detect multiple parallel gray failures~(\cref{sec:eval:robust}).

Specifically, we make the following contributions:

\begin{itemize}[noitemsep,topsep=0pt,parsep=0pt,partopsep=0pt,leftmargin=*]
    \item A passive, coordination-free detector for gray failures in packet spraying networks.
    \item An algorithmic localization system that isolates the failed component to an individual link.
    \item An implementation on Intel Tofino switches with full testbed integration, demonstrated to detect failures during replay of training collectives to show real-world suitability. 
\end{itemize}

This work extends a previous workshop submission~\cite{flowpulse}.

\section{Background and Motivation}\label{sec:background}

The design of \projname is motivated by several trends that have emerged in ML training networks.

\paragraph{ML Training Communication}
Distributed ML training is a highly structured, bulk-synchronous workload.
Nodes alternate between periods of computation and periods of coordinated communication in the form of network collectives.
This structure makes ML training highly susceptible to network delays: a single delayed flow stalls the subsequent computation at the destination node, which rapidly propagates to the entire training cluster during subsequent network collectives.
As a result, even seemingly minor gray failures can have large impacts on end-to-end training performance, as shown in~\cref{fig:allreduce-slowdown}.
The network collectives themselves rapidly saturate the network using only a small number of multi-GiB flows.
This makes it extremely enticing to use flow-level measurements of application traffic to detect gray failures, rather than separate probing traffic.
Not only does this prevent adding additional load to an already-saturated network,
it has only minor overhead in switches since little state is needed to maintain statistics for a small number of flows.
Training clusters commonly use RoCEv2 RDMA as transport protocol. While there are lossy modes for RoCE, lossless versions are preferred~\cite{meta-rdma}. Therefore, every packet drop in the network is considered a network failure. %

\paragraph{Adaptive Routing}
Since ML training traffic is composed of relatively few flows, traditional flow-level load balancing suffers heavily from flow collisions, leading to poor performance~\cite{meta-rdma}.
Instead, many operators have turned to Adaptive Routing (AR), in which switches forward or \emph{spray} packets from a single flow across all available upstream paths towards the destination~\cite{spectrumx-whitepaper, spectrumx-arxiv}. The set of upstream paths is derived from the switch-local routing table. Known failures are represented as missing routes in the table.
Switches may forward packets randomly across 
upstream paths~\cite{random-spraying}, or employ more sophisticated strategies such as selecting the least congested port~\cite{drill}. AR requires a reorder tolerant transport, which is supported for write operations in modern RDMA NICs~\cite{nvidia-ddp, nvidia-cx6-ar}.

AR in non-blocking Clos topologies achieves near-optimal performance with low latency under high demand~\cite{2006-adaptive-routing, drill, cao2013per, power-of-two-choices}.
It has therefore long been the design choice for Infiniband networks~\cite{IB},
and is increasingly deployed in Ethernet backend networks by NVIDIA~\cite{supercharge-ai,spectrumx}, Cisco~\cite{cisco-dlp}, and Broadcom~\cite{BCM}. 
When it comes to network faults, however, AR is a double-edged sword: any flow with a path across a link experiencing a gray failure \emph{will} be affected, so a single failure is likely to affect many flows. This increases the need for reliable gray failure detection in ML networks which use AR.

Because AR techniques prevent end-hosts from learning or influencing the specific path taken by each packet, failure detection in these networks must be conducted using information gathered by switches, instead of by end hosts.

\paragraph{2LFT flat topologies in large-scale ML clusters}
The capacity of modern switches is struggling to meet the demand for high-radix, high-throughput switches. Multi plane topologies consisting of multiple parallel flat networks (planes) with higher radix switches but lower per-port bandwidth are emerging as the dominant solution to the scaling problem~\cite{spectrumx-multiplane-support, oracle-multiplane}. To achieve the same bandwidth as the low-radix topology, operators use multiple, parallel network planes~\cite{multiplane}.
Because of this trend, which allows scaling 2-level topologies to $131k$ GPUs with current switches~\cite{spectrum6800ld}, we focus \projname on flat, \emph{2-level Fat Tree topologies}. %

Despite the smaller size of each plane, the overall network still consists of a large number of links and switches, meaning that components are frequently taken down for maintenance or due to previously detected failures. The asymmetry in such networks is a steady state, not an exception~\cite{kokolis2025revisiting}.

\paragraph{Opportunity: Predictability of Adaptive Routing}\label{sec:predictability}
There are many sources of non-determinism in networked systems, making it impossible to predict the path of an individual packet in AR networks.
However, when aggregating over many packets, a predictable \emph{spraying distribution} emerges which can be modeled.
For a symmetric 2-level fat tree topology without failures, the symmetry of the network results in even spraying in expectation. In other words, an equal number of packets are expected to be received at the destination leaf switch from each spine.

\begin{figure}
    \centering
    \includegraphics[width=0.8\linewidth]{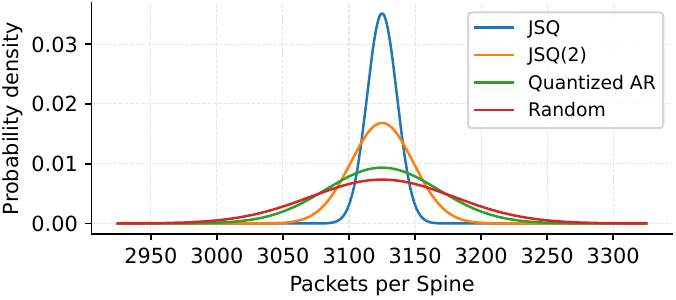}
    \vspace{-1em}
    \caption{Distributions for AR spraying strategies: JSQ: Join Shortest Queue, JSQ(2): Power of Two Choices~\cite{power-of-two-choices}, Quantized Adaptive Routing~\cite{loadbalancing-for-ai}, and random. $100K$ packet flow sprayed across $32$ spines. $\mu=\frac{100k}{32}=3125$\simulation}
    \label{fig:loadbalancing}
\end{figure}

In practice, some variance from this ideal distribution should be expected.
\Cref{fig:loadbalancing} shows observed packet spraying behavior for several different AR policies. "Random" sprays each packet via a random spine, while Joint Shortest Queue (JSQ) always sprays via the shortest queue. In-between, JSQ(2) selects two random queues and sprays via the shorter, also known as "power of two choices"~\cite{power-of-two-choices}, while Quantized AR sorts queues into buckets based on their current length and selects a queue from the shortest bucket~\cite{loadbalancing-for-ai}. For all policies, the observed packets per spine follows a Gaussian distribution around the expected value for even spraying.
As more packets are sampled, the Gaussian distribution becomes tighter around the expectation.

When a gray failure appears in the network, some packets will be dropped along paths containing the failure.
This results in fewer packets than expected arriving at the destination leaf switch from the corresponding spine.
Once enough packets are sampled, standard statistical hypothesis tests can be used to determine if a given spine is a part of a healthy path, or if a gray failure is causing packet drops.

\noindent\emph{Takeaway: AR results in a predictable, even spraying distribution for all analyzed load balancing strategies. If the destination leaf switch observes fewer packets from one spine, this implies the presence of a gray failure.}

\section{Design}\label{sec:design}
\projname is a passive, low-overhead, in-network gray failure detector that predicts the AR packet spraying distribution and compares it against runtime observations. It is applicable to networks using AR~\cite{spectrumx-arxiv} with large (multi GB) individual flows between leaf pairs. This requirement is satisfied for AR-based AI training clusters.
To illustrate the reasoning behind the design, we first present a strawman for pristine, symmetric networks. We then show how network asymmetries break this approach, present the full \projname design, and elaborate the core components in detail.

\subsection{Strawman Approach}
In a symmetric network, where all links are healthy and available for AR, the source leaf switch sprays a flow evenly across all spines (as discussed in~\Cref{sec:predictability}). In turn, the destination leaf switch receives the same number of flow packets from each spine.
A gray failure on a single link reduces the number of packets the destination leaf receives from the spine connected to the failed link as the packets are dropped along the way. 
The failure detector can therefore count how many packets of the flow it receives from each upstream spine and check if it received fewer packets from one spine than from others. If so, it flags that upstream path as failed.

While this approach works in symmetric networks, in reality, %
networks are asymmetric, either due to preexisting failures or maintenance. This results in an uneven spraying distribution, breaking this approach as we illustrate next. Pure load imbalance can no longer serve as a failure signal.

\subsection{Asymmetry in AR \& Prioritization}\label{sec:asymmetry-predictability}

\begin{figure}
    \centering
    \subfigure[Flows A and B\newline fully overlap.\label{fig:asymmetric-topology:same-runtime}]{
    \includegraphics[width=0.30\linewidth]{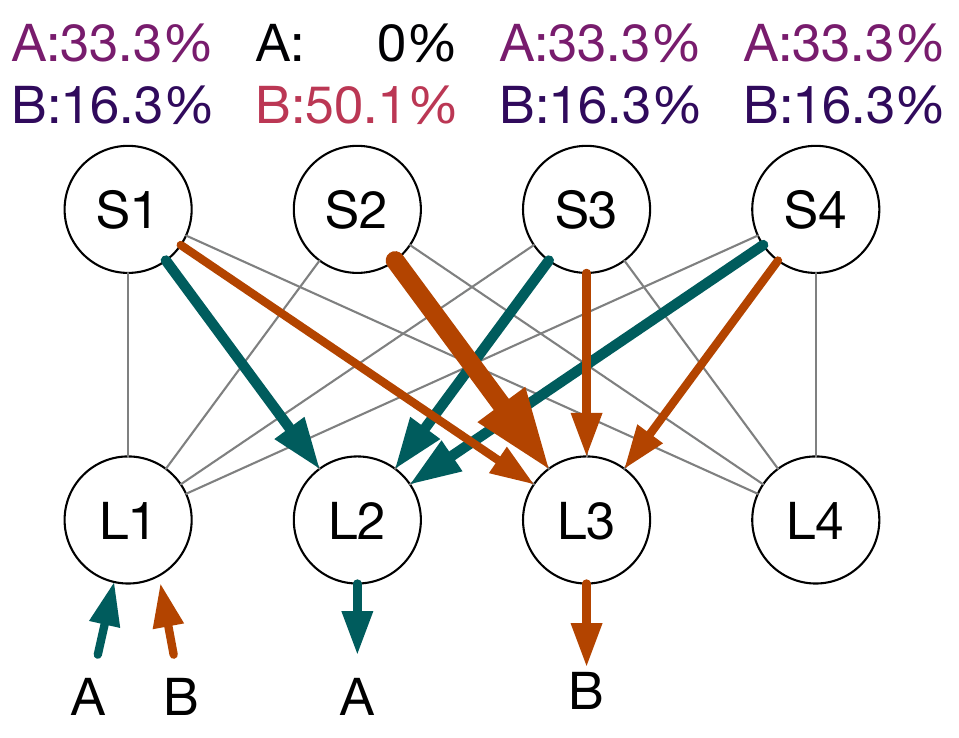}
    }
    \subfigure[Flow A partially\newline overlaps with flow B.\label{fig:asymmetric-topology:timing}]{
    \includegraphics[width=.30\linewidth]{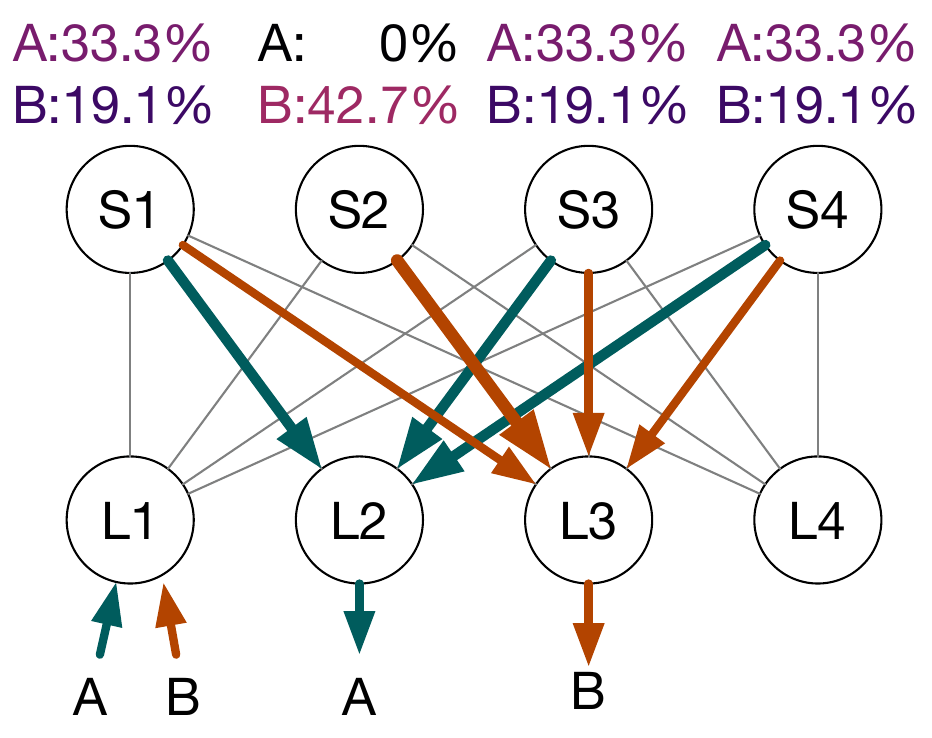}
    }
    \subfigure[Flow B prioritized.\label{fig:asymmetric-topology:prio}]{
    \includegraphics[width=.30\linewidth]{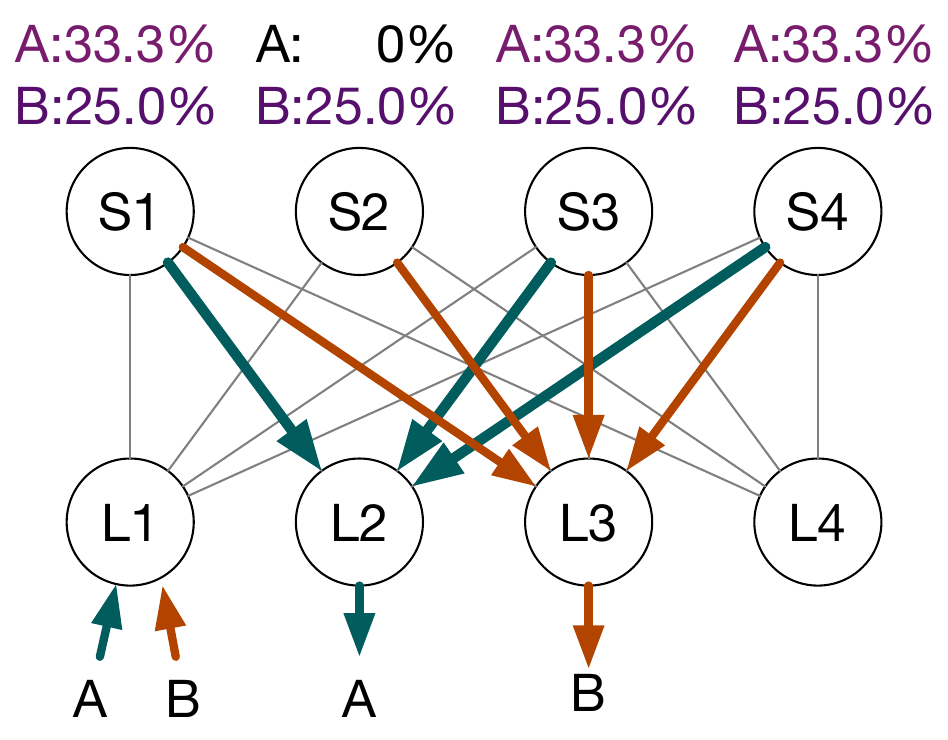}
    }
    \vspace{-1em}
    \caption{Asymmetric topology using JSQ(2) spraying. Flow A is sprayed across spines 1,3,4 while flow B is sprayed over all spines. Flow B's spraying distribution depends on its timing relative to flow A unless it is prioritized.\simulation}\label{fig:asymmetric-topology}
    \vspace{-1em}
\end{figure}

Asymmetric topologies with competing flows make it hard to predict a flow's spraying distribution when the flows only partially share the upstream links used for packet spraying.

\Cref{fig:asymmetric-topology} shows three scenarios of two flows being sprayed from the same source (leaf L1) to different destination leaves. Both flows are the same size and are sprayed across partially overlapping sets of upstream paths.
\Cref{fig:asymmetric-topology:same-runtime} shows that when both flows run at the same time, A is sprayed nearly equally, but flow B, which can use spine S2 exclusively, sends the majority of the traffic through S2. However, in~\cref{fig:asymmetric-topology:timing}, when starting flow B before A, flow B is sprayed more balanced, since it uses the entire network exclusively before flow A starts. 
We conclude that the packet spraying distribution in asymmetric networks with competing flows depends on the \emph{relative timing} between the flows and is therefore largely unpredictable without knowing this timing. The strawman approach no longer works.

There is however a way to restore equal spraying for flow B: \emph{network prioritization}.
Network switches implement different priority levels. Commonly, eight priority levels can be used for quality of service~\cite{8021Q}. They are implemented as separate egress queue per priority level on each port.
AR sprays the packets over the set of queues from the same priority level based on the aggregate queue length of all higher priorities. Meaning that if there is only a single flow with the highest priority during spraying, it will not be affected by competing flows and will spray packets as if it is the only flow in the network. Lower priority flows, however, are sprayed based on the queue lengths of the higher priority levels. 
\Cref{fig:asymmetric-topology:prio} shows the distribution with flow B prioritized. It now is equally sprayed across all paths. Flow A also sprays equally in this case as all higher-priority queues are of the same length because B uses all paths in the network.  

Note that the actual load on the whole switch stays the same as before, keeping the conditions in which the flows pass through it intact, and only affecting the spraying distribution. 

\noindent\emph{
Takeaway: In an asymmetric network, if a single flow is given the highest priority,
the spraying pattern of that flow is not affected by concurrent traffic.
}

\projname resolves the challenge of competing traffic in asymmetric topologies by prioritizing a single cross-leaf measurement flow on each leaf switch (\Cref{sec:design:flow-selection}).
\projname then checks for network failures by testing runtime observations of how the prioritized flow is sprayed in the network against the analytical prediction on the destination leaf. For this, it calculates a detection threshold based on the flow information and its local routing table (\Cref{sec:design:threshold}). When the flow finishes, the switch compares its runtime counters.
If a spraying path sees fewer packets than can be explained solely by the distribution of the load balancer, \projname concludes that there must be a gray failure (\Cref{sec:design:detection}). The selection of prioritized flows, performed by the source switches, changes over time to achieve high coverage of the fabric links.

\subsection{\projname Overview}
\begin{figure}
    \centering
    \includegraphics[width=\linewidth]{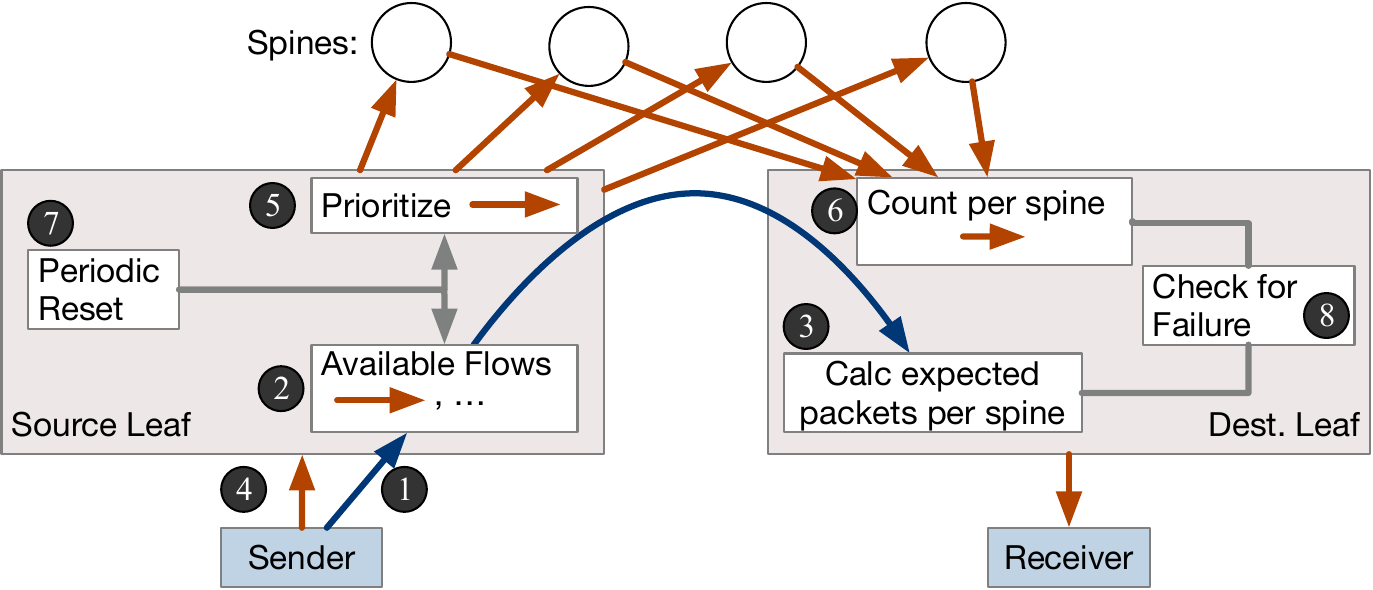}
    \caption{\projname Overview and Walkthrough.}
    \label{fig:design-overview}
\end{figure}

We now provide an overview of \projname's systems components and their interactions as shown in~\cref{fig:design-overview}.

\noindent\textbf{End Hosts:} At the start of a collective, the collective library 
sends a \emph{flow announcement packet} \circled{1} to the flow's destination. This packet contains the flow size and the destination queue pair (QP)  
numbers, which allows the source leaf to know about all flows it can select from for prioritization and the destination leaf to know the flow size and thereby expected number of packets.
The packet's overhead is negligible compared to the flow size (17 byte announcement per flow).
The host starts the flow afterwards~\circled{4}.

\noindent\textbf{Source Leaf:} Upon receiving the announcement, the source leaf marks the flow destination as available for its flow selection policy \circled{2}. It selects one cross-leaf flow at a time (\cref{sec:design:flow-selection}), prioritizing its packets \emph{in that leaf only} to isolate the spraying behavior \circled{5}. Once the flow completes, the source leaf marks the destination switch as covered. The flow tracking and coverage map are periodically reset \circled{7} to avoid stale information.

\noindent\textbf{Destination Leaf:} This switch performs the core detection logic. Using the metadata from the announcement and local routing tables, it predicts the flow's spraying behavior \circled{3} and calculates a failure detection threshold (\cref{sec:design:threshold}). As the flow runs, the switch counts the packets of that flow as received from each upstream port \circled{6}. After the flow finishes  the switch compares these measured counts to the threshold \circled{8} and alerts the monitoring system if a failure is detected (\cref{sec:design:detection}). If more packets are needed to reach statistical significance, the measurements extend to additional flows.

We now discuss our design in detail.

\subsection{Flow Selection and Isolation}\label{sec:design:flow-selection}
The source leaf selects a flow which it was informed of using the flow announcement packet.
Each leaf switch selects a single outgoing (spine-bound) flow at a time for measurement, and prioritizes the packets of that flow at the highest priority in that switch (which is reserved exclusively for this purpose) to isolate its spraying distribution.
Since prioritization is only needed to isolate spraying, it is only performed at the source leaf switch.
Otherwise, all switches honor user-defined priority levels.

A key question is how to select flows in a way that maximizes the coverage of the fabric links.
\projname achieves this using a coordination-free flow selection strategy.

Under \projname, each source leaf uses a \emph{local round robin (RR)} schedule across possible destination leaves to select the measurement destination.
Once a given leaf is selected, the next flow destined to that leaf is selected for prioritization.
This ensures that all used network paths reachable from this switch are covered, as packet spraying ensures that all the paths between every two leaves are explored. 

Because \projname uses application packets for measurement, it can only cover a given destination leaf if a flow is actually sent to it.
To avoid blocking indefinitely on flows that never arrive, leaf switches keep track of which destinations are used by the current application, and only include those destinations in the RR.
\projname accomplishes this by maintaining a bitmap of currently-reachable destinations at each leaf switch.
When the switch observes a flow announcement packet, it marks the destination as reachable.
To account for changing network traffic patterns, the control plane regularly resets the bitmap, ensuring that the flow selector does not get stuck due to stale information.

\subsection{Spraying Prediction}\label{sec:design:threshold}
\projname detects gray failures by predicting the expected load balancing distribution of a flow and comparing runtime measurements to this expectation.
Measurements that deviate sufficiently from this expected distribution indicate the presence of a gray failure. 
We now discuss how \projname predicts the spraying distribution and calculates the detection threshold.

\paragraph{Expected Per-Spine Load}
Consider a single flow of $N$ packets from a source leaf switch to destination leaf switch, routed through $k$ spine switches.
Packets are sprayed across all $k$ candidate spines, and each leaf-spine link carries, in expectation, $\lambda = \mathbb{E}[X_i] = N/k$ packets.
Here, $X_i$ is a random variable representing the observed number of packets arriving from a given spine at the leaf switch, assuming no gray failures.
Both $N$ and $k$ are known to the control plane: the flow size $N$ is learned from the flow announcement, and $k$ is the number of usable paths according to the local routing table.

The variance of $X_i$ around $\lambda$ depends on the spraying policy.
For purely random uniform spraying, each packet independently selects a spine, and $X_i$ follows a binomial distribution with
$\sigma^2 = \text{Var}[X_i] \;\approx\; \lambda$. %
Other policies are instead driven by queue occupancy, which \emph{tightens} the distribution, lowering variance compared to the random baseline as shown in~\cref{fig:loadbalancing}.
Regardless of the spraying policy, for a large number of packets sent during the same ML collective ($>10^5$ per flow), the distribution of $X_i$ is normal due to the central limit theorem.

\paragraph{Effect of a Gray Failure}
A gray failure on the path via spine $j$ that drops packets at rate $p$ reduces the expected count from that spine to $\mathbb{E}[X_i \mid \text{failure}] \;=\; \lambda \cdot (1 - p)$,
producing a \emph{deficit} of $p\lambda$ packets relative to the expected load.
This deficit allows us to detect gray failures using standard statistical methods.
Specifically, we use a one-sided Z-test, a standard hypothesis test for normal distributions.
Our null hypothesis is a healthy path with a mean of $\lambda$, while our tested hypothesis is a lower mean due to a gray failure.

\paragraph{Threshold Selection}
Following the Z-test, we flag a failure on the path via a given spine $i$ whenever $X_i$, the number of packets observed arriving from spine $i$, is below a threshold $t = \lambda - s \sqrt{N/k}$, where the parameter $s$ determines the sensitivity. $s$ can be chosen either analytically based on $\sigma^2$ and the desired detection confidence, or by empirically checking which value results in the desired confidence on a given network (we use the latter approach for our evaluation; see \cref{sec:eval:detection}).
Because $\lambda$, $s$, $N$, and $k$ are known to the switch at the beginning of each flow, the detection threshold can be determined immediately without the need for live network observations.

\paragraph{Effect of Flow Size}
The detectability of a failure having drop rate $p$ depends on the ratio of the packet deficit on a spine to the noise of the load balancing algorithm.
The deficit is $p\lambda = pN/k$, while for random spraying, the noise scales as $\sigma \sim \sqrt{\lambda} = \sqrt{N/k}$, resulting in a signal-to-noise ratio (SNR) of 
$\frac{p\lambda}{\sigma} \;\approx\; p\,\sqrt{N/k}$.
The SNR improves based on the number of \emph{packets per spine}. In effect, larger flows produce tighter \emph{relative} distributions around~$\lambda$, making even small drop rates detectable.
We quantify this relationship empirically in~\cref{sec:eval:detection}.

\paragraph{Cross-flow Aggregation}
When used to detect gray failures with very low drop rates, or on networks with many spines, a single
flow
may not send a sufficient number of packets for \projname to achieve high accuracy.
In this case, data can be aggregated across multiple flows between the same source and destination leaf switches, for example across multiple collective iterations.
This increases the number of packets $N$ used for detection, strengthening the measurement signal and improving sensitivity. Due to the low number of flows (\cref{sec:background}), the required in-switch state is small enough to retain flow statistics for aggregation within the switch's memory limitations.

\subsection{Failure Detection and Localization} \label{sec:design:detection}

After
a leaf switch
sees the last packet of a measured flow, identified by the maximum expected sequence number, it compares the per-spine packet counters to the detection threshold calculated for that flow.
If it received too few packets from a given spine,
then it notifies the network monitoring system that a failure has been detected on the path via that spine.

\begin{figure}
    \centering
    \includegraphics[width=0.6\linewidth]{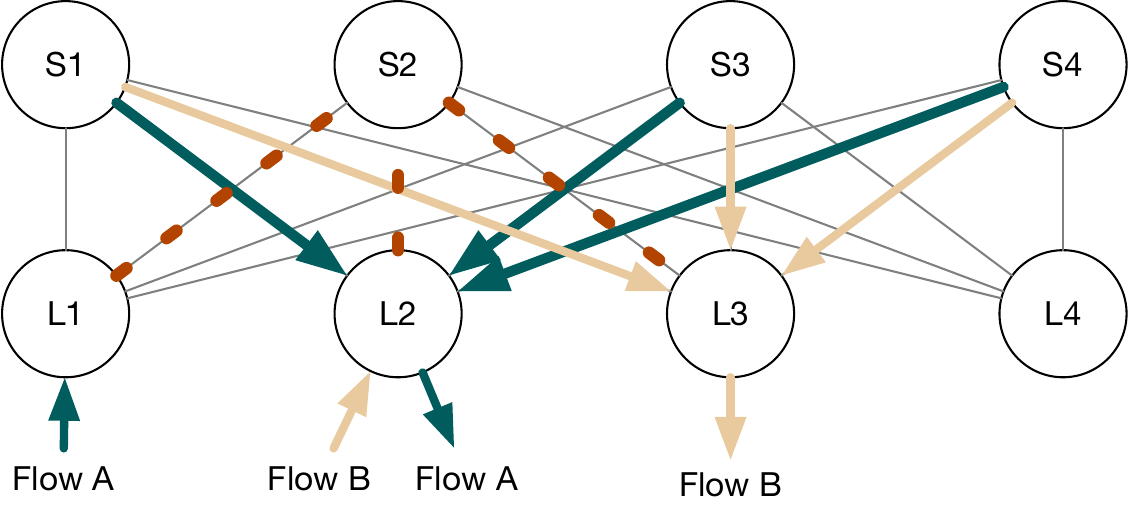}
    \caption{Example Localization Mechanism. The link between \texttt{L2} and \texttt{S2}, notated as \texttt{L2S2}, failed.
    Measurement flow A and B detect the links $\lbrace\texttt{L1S2, L2S2}\rbrace$ and $\lbrace\texttt{L3S2, L2S2}\rbrace$ as failed, respectively. The localization algorithm takes the intersection of all failure reports involving \texttt{S2} and concludes that the link \texttt{L2S2} failed.}
    \label{fig:localization-example}
\end{figure}

\projname localizes failures in the central monitoring system, which receives all failure notifications.
While leaf switches may be able to localize some failures on their own, centralized localization has several advantages.
On-switch localization is both workload-dependent and vulnerable to multiple failures shadowing each other, neither of which is a problem for centralized localization.
At the same time, the monitoring system must be informed of failures regardless of where they are localized.
Given the rarity of gray failures, as long as \projname is configured to achieve high detection precision, the overhead of path-failure notifications is small.

When the central monitor receives a failure report,
it flags the entire path between the source and destination switch as potentially failed. This path consists of two links: the uplink from the source leaf to the spine (link 1), and the downlink from the spine to the destination leaf (link 2).
To determine which has failed, the central monitor waits for failure indications from other flows. If link 1 has failed, then additional flows from the same source leaf to different destination leaves will report a failed path including link 1, and similar for link 2 and flows from different source leaves.
In short, a link is considered failed when it is in the intersection of multiple failure reports that include a different leaf switch.

\Cref{fig:localization-example} shows an example where the two flows, from \texttt{L1} to \texttt{L2} and from \texttt{L2} to \texttt{L3}, enable a failure to be localized. In this case, the link between \texttt{L2} and \texttt{S2} is failed. When checking all reports involving \texttt{L2}, the localization algorithm correctly concludes that \texttt{L2S2} failed.
Since flows not involving \texttt{L2} can still spray via \texttt{S2} to \texttt{L1} and \texttt{L3}, the algorithm will not conclude that \texttt{L1S2} or \texttt{L3S2} have failed.

\paragraph{Multiple Failures}
Our localization algorithm is robust to situations with multiple parallel failures, localizing all failures individually. There are three possible ways for failed links to be positioned relative to each other: (1) two failed links sharing the same spine (and connected to two different "victim" leaf switches), (2) two failed links sharing the same leaf, and (3) two failed links sharing neither leaf nor spine. In both the second and third case, both failures appear in disjoint sets of paths; since our algorithm localizes failures using path intersections, both failures will be localized independently.
While the first case poses a risk of both failures shadowing each other, as long as there exist two flows which each involve a different victim leaf switch
(not counting flows which involve both simultaneously),
our algorithm will be able to identify both failures.

\paragraph{Traffic Assumption}
As \projname targets AR networks, the localization algorithms assumes that traffic is sprayed across all spines and therefore, all upstream links are validated. \projname cannot detect failures on links unused by the application traffic. 
Further, if the there is only communication between two fixed leaves, \projname cannot isolate whether a failure is on the up- or downlink of a spine. However, these traffic patterns are uncommon in ML training clusters where AR is commonly deployed.

\section{Implementation}\label{sec:implementation}
To show the feasibility of in-network failure detection using load imbalance as failure signal, we prototype \projname using Intel P4 Studio~\cite{p4-studio} and Tofino-1 programmable switches~\cite{tofino}. 
This section presents each component's implementation. While we divide source and destination leaf for clarity, every leaf always fulfills both roles. \Cref{appx:routing} describes the assumed routing system.

We extend UCC~\cite{ucc}, the collective communication library (CCL) of UCX~\cite{ucx}, to send the flow announcement packet containing the flow identification and size.
We believe that this functionality can be implemented in any other CCL. 
The implementation is oblivious to specific collective algorithms, as it only requires network flow information and is limited to an addition of $133$ LoC in the UCC profiling subsystem to extract the flow size and send the packet to the flow's destination.

\subsection{Source Switch}
The source switch data plane selects the measurement flow and isolates its spraying from competing flows.

\paragraph{Flow Selection}\label{sec:implementation:flow-selection}
Each leaf selects exactly one cross-leaf flow for measurement. %
To select a flow, the switch data plane parses the flow announcement packets to track which flows are available in the system, and maintains a bitmap of the destination switches to which flows are available.
To maintain the round-robin policy of flow selection, the switch keeps another bit mask containing the history of destination switches it already covered. It then selects the flow to the lowest destination switch index which has not been recently selected and to which a flow is available. To ensure progress, the switch control plane resets the history bit mask and available flow mask regularly, e.g.,  every minute. 
After selecting the prioritized flow, the switch marks all packets of the flow as measurable for \projname using the lowest bit in the first reserved block of the RoCE base transport header.

\paragraph{Spraying Isolation}
When the flow is selected by the selection policy, the switch assigns it to the highest priority queue in the sending leaf switch. All packets arriving with user-defined priority levels below the maximum are assigned according to the defined level. We reserve the exclusive use of the highest priority in the network for \projname. Users may use all other priority levels. The exclusive usage ensures that \projname does not change the user-defined priority order. No other switches in the network prioritize the measurement flows.
Limiting the prioritization to the spraying switch minimizes the performance impact of using network priorities. 

Since the prioritized flow is sprayed across all upstream ports, the fraction it occupies on a single port is relatively small compared to the port's overall capacity.
If a prioritized flow is sprayed across 64 ports, only $1.56\%$ of each port are used by the prioritized flow, leaving the remaining capacity for other traffic. We evaluate the impact in~\cref{sec:evaluation:assumptions}.

\subsection{Destination Switch}
The destination switch implements the detection mechanism of \projname. It computes the detection threshold, counts the prioritized packets received per spine from each flow, compares the counters to the threshold, and alerts the network monitoring system when a failure is detected.

\paragraph{Threshold Calculation}
After receiving the flow announcement packet, the data plane parses the packet contents into switch memory for the control plane to read and calculate the threshold according to~\cref{sec:design:threshold}.
The threshold calculation cannot be implemented in the data plane because of the complex arithmetic operations required. We therefore resort to the control plane. We avoid packet parsing in the control plane because of security concerns.
The control plane stores the per-spine threshold and the maximum expected packet sequence number (PSN) in data plane registers, indexed by the destination queue pair (QP). 
To avoid stale information, the control plane keeps a queue of QP numbers and erases their data plane state after a timeout of $1$ minute, assuming that no flow will take longer to complete.

\paragraph{Packet Counting}
The data plane counts how many packets marked as measurable it received from each source leaf via each spine. The switch keeps one 16-bit counter for each, accumulating to less than 2KB of memory in a 32-spine topology. The packets must be marked as measurable by the source leaf to be counted. While the marking does not actually prioritize the packets, it allows the switch to specifically count only source-prioritized flows. 
The switch counts all marked packets, regardless of if the threshold value is available in memory or not. This is necessary in order to account for reordering of the announcement packet and the threshold computation time.
It stores the packet counts in a map indexed by the destination QP number and spine switch from which the packet was received. This allows the simultaneous measurement of flows from different sending leaves. The switch also stores the expected highest PSN as calculated from the flow size. When it sees this PSN, it stops counting and performs the failure detection.

\paragraph{Detection}
To detect failures, the switch compares the received per-spine packet number with the calculated threshold. If the counter is lower than the threshold it marks the packet path as failed in a failed-paths bitmap. The control plane regularly reads the bitmap and notifies the network monitoring system about the detected failure.

\paragraph{Sensitivity to packet reordering}
The detection of the flow's last packet may not be accurate due to packet reordering.
Packet spraying can result in reordering as the delays of different paths may not be identical.
We show in~\cref{eval:reordering} that a realistic amount of reordering is well below the level that may impact \projname's detection accuracy.

\paragraph{PFC}
When a switch receives a PFC pause frame, it stops sending to the port on which the packet is received until it receives a resume frame. If the leaf receives a pause frame on a port, it does not spray packets via this upstream port. PFC thereby alters the packet spraying distribution. To avoid false positive failure detection, \projname discards measurements during which PFC events were detected.

\section{Evaluation}\label{sec:evaluation}
We evaluate \projname's detection quality, exhibiting how the accuracy depends on the network and workload, and its robustness to congestion control and concurrent network load.

\subsection{Setup}
We evaluate \projname both in a real-world testbed and with packet simulations.
For small-scale experiments, we use our testbed, which emulates a full two-level fat tree topology with up to 8 spine and 8 leaf switches. We only simulate a single host per leaf, as \projname only uses cross-leaf traffic for failure detection.
We use NS-3~\cite{ns3} network simulations to evaluate \projname in larger topologies. We confirm that the results from the testbed and from the simulation for the same-scale topologies are on par.

\begin{figure}
    \centering
    \includegraphics[width=1\linewidth]{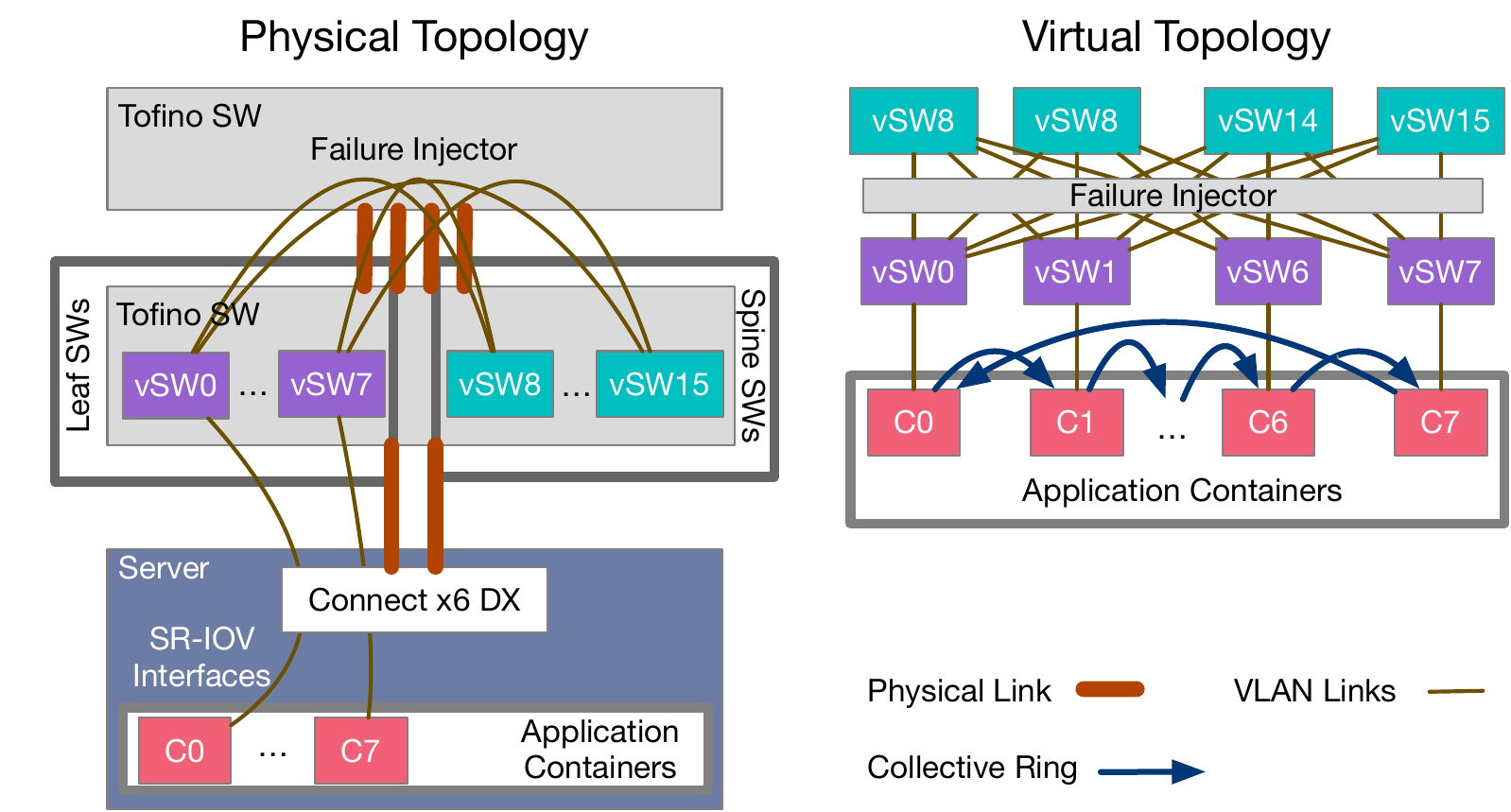}
    \caption{Testbed Architecture. The virtual topology is implemented by multiplexing the Tofino switch into 16 virtual switches connected via VLANs. \projname is implemented in the leaf vSwitches.}
    \label{fig:testbed-overview}
\end{figure}

\paragraph{Testbed}
The left half of \Cref{fig:testbed-overview} depicts the physical topology of the testbed, which consists of
two Intel Tofino-1 switches, one running multiple virtual switches and one injecting failures. We use the virtual switches to build the virtual topology shown in the right half of \Cref{fig:testbed-overview}, a non-blocking two level fat tree with 8 leaves and 8 spines. This is the largest topology we could fit in a Tofino-1 switch.
Each virtual link is implemented as an isolated VLAN on four shared physical links.
Traffic is generated by containers running on a Linux server connected with a NVIDIA ConnectX-6 DX~\cite{connectx6-dx} NIC. The NIC exposes SR-IOV virtual functions (vNICs), one for each container.
All vNICs use RDMA with DCQCN as congestion control, and support selective repeat loss recovery and out-of-order packet delivery, specifically enabled for these NICs by NVIDIA. 
As the testbed only serves for simulating the effects of spraying in a network, rather than absolute performance measurements, we believe it serves as a reliable proxy for estimating the behavior of a real system.
We further describe the testbed setup and implementation in \cref{appdx:testbed}.

\paragraph{Packet Simulation}
We use NS-3~\cite{ns3} for network simulation with the Astra-SIM RDMA implementation~\cite{astra-v2}. We simulate non-blocking 2-level fat tree topologies connected with $100$Gb/s links. The topology is always full, so its size is determined by the number of spines, and denoted for each experiment.
We extended the simulator to support out-of-order packet delivery and selective repeat (SR) loss recovery~\cite{selective-repeat}. As in the testbed we use DCQCN~\cite{dcqcn} for congestion control. We use JSQ(2) (join shortest queue with random tie breaking) per-packet adaptive routing on all leaf switches. We use MPI with UCC to generate the collective traffic across the 8 ranks. Unless otherwise mentioned, we do not generate additional background traffic on the network.
We run each simulation $20$ times with different seeds for randomization.

\paragraph{Workload} In both the testbed and simulation, we use a \ringallreduce workload of 2 GiB across 2 Queue Pairs per rank unless mentioned otherwise, resulting in a flow size of 1 GiB.
We run one endpoint on each leaf switch; since \projname's failure detection is based on spraying distributions across spines, local traffic within a leaf is not considered and we omit it during experiments.

\paragraph{Failure Injection}
We inject failures by randomly dropping packets with a certain probability on a single fabric link. We do not inject failures on access links, i.e., links between the switches and end hosts.

\subsection{End-to-end Evaluation}
\begin{figure}
    \centering
    \includegraphics[width=0.8\linewidth]{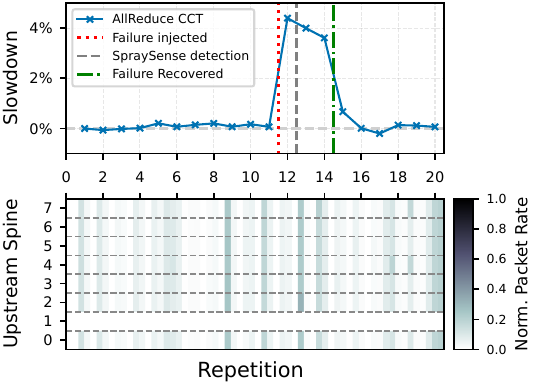}
    \vspace{-1em}
    \caption{Failure detection when running 20 repetitions of \allreduce. 
    Packet drop rate of $1\%$ is injected on a single link before repetition $12$.
    Top:  Relative slowdown of the collective completion time.
    Bottom: Packet arrival rate at each upstream port. The link to spine \#1 is deliberately disabled.
    \projname detects the failure immediately after the end of repetition $12$. The failure is not visible via packet rate telemetry. \testbed.}
    \label{fig:eval-works}
\end{figure}

We demonstrate the end-to-end use of \projname with the following testbed experiment in \Cref{fig:eval-works}.
We run $20$ repetitions of \ringallreduce, and inject a failure on a single link by setting its drop rate to $1\%$. In addition, the fabric is asymmetric with two links permanently taken down in the routing tables, one from the source leaf to spine 4, and from the measurement leaf to spine 1 (this link is visible as the empty row in the lower graph since it is fully disabled).
In addition to each collective, we run a bisection flow at line-rate to the measurement switch to create background traffic. The bisection flow's sender is placed under a different leaf and can use paths via all spines except spine 1 due to the disabled down link. %

\Cref{fig:eval-works} shows a failure injection before repetition 12 and \projname detecting it immediately after the repetition. The injected failure causes a slowdown of the collective. During the failure, the per-port packet rates, measured from the destination leaf's data plane, do not exhibit any distinctive change. In contrast, \projname successfully identifies the fault.

\subsection{Calibration of Sensitivity and Accuracy}
\label{sec:eval:detection}\label{sec:evaluation:detection}

Our goal is to detect failures that might induce slowdown on the performance of collectives as early as possible, while at the same time
avoiding both false positives and false negatives.
Setting a tight detection threshold (\cref{sec:design:threshold}) increases the sensitivity to a lower drop rate, but can also decrease detection accuracy due to inherent spraying variance. At the same time, to reduce the noise, more packets need to be counted in each measuring switch to achieve statistically significant results, thus potentially affecting more collectives longer. Decreasing the sensitivity with a more relaxed threshold would imply more significant degradation of a single iteration of a collective but more rapid and robust detection. 

Conceptually, we prefer to detect failures quickly and with the highest robustness. Thus, we seek to find the lowest drop rate with minimal effects on the system performance, which affords perfect detection accuracy, and can achieve this with a realistic number of packets to allow fast detection.  

More formally, we need to set two parameters of the detector at deployment time: sensitivity $s$, which determines the percentage of drops per link we detect with perfect precision, and the minimum number of packets $P_{min}$ per flow per spine port necessary to count until reaching statistical significance.

The calibration requires optimizing a Pareto frontier over $s$ and $P_{min}$ to find their values that allow detecting a certain per-link failure rate with perfect accuracy of 100\% TPR and 0\% FPR.
We take a simplified iterative approach instead: we first calibrate $s$ with a large number of packets per spine, and then reduce the number of packets to find $P_{min}$ given $s$. 

To find $s$, we run the detector in the testbed with various injected drop rates, and record the respective packet distributions. We use these to find the value for $s$ that results in a perfect detection of the lowest possible drop rate by calculating the respective Receiver Operating Characteristic (ROC) curves. Now, given $s$, we run the detection with varying number of packets per spine to find $P_{min}$, reevaluating the ROC to ensure perfect detection accuracy.
\begin{figure}[t]
    \centering
    \subfigure[\simulation]{
    \includegraphics[width=0.47\linewidth]{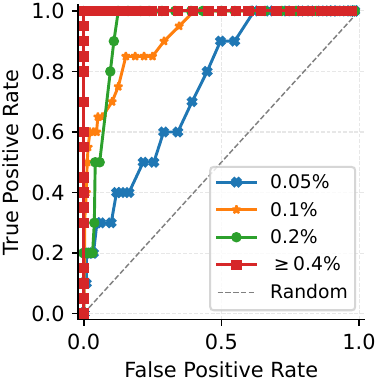}
    }\hfill
    \subfigure[\testbed]{
    \includegraphics[width=0.47\linewidth]{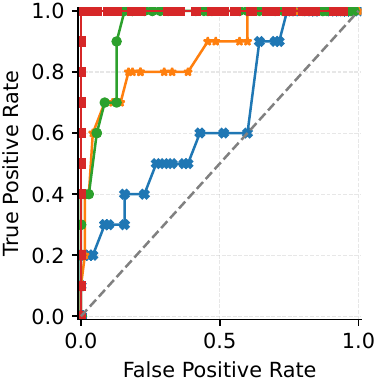}
    }
    \vspace{-1em}
    \caption{ROC Curves for different values of $s$. \projname perfectly detects loss rate of $\geq0.4\%$ on a single link in an 8 spine topology, both in simulation and on the testbed with a $500k$ packet measurement flow size}
    \label{fig:roc-curves}
\end{figure}

\paragraph{Calibration on the testbed}
\Cref{fig:roc-curves} presents ROC curves to find $s$. 
\projname achieves perfect accuracy for drop rates $\geq0.4\%$ on a single link in the 8-spine topology with $500K$ packets per spine (\Cref{fig:num-packets}). We achieve zero FNR and zero FPR (not shown) by setting $P_{min}$ to 60k for detecting 0.5\% per-link drop rate, 20k  for 1\% and 7k for 1.5\%. 

\paragraph{Simulation vs. Testbed}
The figure depicts the ROC curve produced by the same setup in the testbed and in the simulation. 
The slightly better results of the simulation compared to the testbed in~\cref{fig:roc-curves} are caused by the approximate implementation of the JSQ(2) load balancing in the testbed, which is more noisy than the exact queuing implementation of the simulation.
However, \emph{sensitivity values for perfect detection accuracy are the same for both.}  
This confirms that we can use simulation to extrapolate the results to larger topologies while using data obtained from the testbed.

\begin{figure}[t]
    \centering
    \subfigure[Finding $P_{min}$ empirically in the testbed \label{fig:num-packets}\testbed]{
        \includegraphics[width=0.45\linewidth]{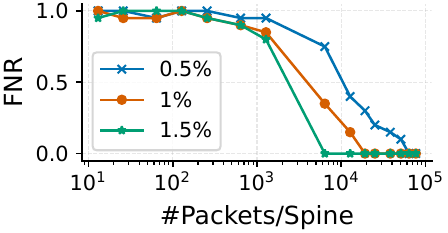}
    }
    \subfigure[Precision for larger topologies with calibrated $P_{min}$ and $s$\label{fig:spine}\simulation]{
        \includegraphics[width=0.45\linewidth]{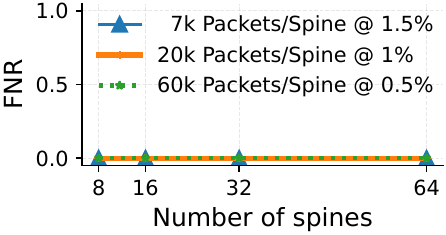}
    }
    \vspace{-1em}
    \caption{Finding $P_{min}$ and validating its precision across multiple topology sizes. (b) The FNR is 0 in all measurements.}
    \label{fig:num-packets-and-switch-radix}
\end{figure}

\paragraph{Extrapolating to larger topologies}
To validate that these results hold for larger topologies, we run a simulation while using the parameters we found in the testbed, and measuring the resulting detection accuracy. 
\Cref{fig:num-packets-and-switch-radix} shows that we achieve the \emph{same} accuracy across different number of spines.

\paragraph{Time-to-detection in the context of model training}
To understand the range of acceptable values for $P_{min}$, recall that this number represents the portion of the packets received at the measuring switch from a certain flow that are forwarded through a single spine. Thus, from the user perspective, $P_{min}\cdot{}N_{spines}$ is the number of packets that the application must send between the same source and destination switch to allow \projname to successfully detect a fault on the path between them.  Therefore, large topologies need proportionally more packets for detection. In the context of collectives, if this number is too high then more collectives are affected until the detection occurs, with higher application-visible impact. 

To understand the application impact of this detection delay, we compare the required $P_{min}$ for different topology sizes with the data sent by a single GPU during a training iteration in its \allreduce collectives. The collective sizes are based on training a Llama-3 70B model with a typical training configuration (4TP/4PP/4DP, 16 $\mu$batches, batch size 256)~\cite{llama3-herd}. 

\Cref{tab:required-collective-sizes} shows the results.  We observe that our values of $P_{min}$ are practical for large topologies and high sensitivity values. For example,  even the highest sensitivity of 0.5\% dropped packets per link allows \projname to detect a problem within 5 training iterations in a large 64-spine topology.

\begin{table}
    \centering\small 
    \begin{tabular}{c|c|cccc}
    \toprule
    \head{1cm}{Loss Rate} & \head{1cm}{$\frac{k Packets}{Spine}$}& Spines &  $k$Packets &\head{1.25cm}{Flow Size [GiB]} & Iter.\\
\midrule
\multirow{3}{1cm}{$2.0\%$}  & \multirow{3}{1cm}{$2$} & 32 & $64$ & $0.56$ & $0.07$\\
& & 64 & $128$ & $1.12$& $0.15$\\
& & 128 & $256$ & $2.23$ & $0.29$\\
    \midrule
\multirow{3}{1cm}{$1.5\%$} & \multirow{3}{1cm}{$7$} & 32 & $224$ & $1.95$ & $0.26$\\
& & 64 & $448$& $3.91$& $0.51$\\
& & 128 &$896$ & $7.81$ & $1.02$\\
    \midrule
\multirow{3}{1cm}{$1.0\%$} & \multirow{3}{1cm}{$20$} & 32 & $640$& $5.58$& $0.73$\\
& & 64 &$1,280$ & $11.16$& $1.46$\\
& & 128 & $2,560$& $22.32$& $2.93$\\
    \midrule
\multirow{3}{1cm}{$0.5\%$} &\multirow{3}{1cm}{$60$} & 32 &$1,920$ & $16.74$ & $2.19$\\
& & 64 & $3,840$& $33.48$& $4.39$\\
& & 128 &$7,680$ & $66.96$& $8.78$\\
    \bottomrule
    \end{tabular}
    \caption{Measurement collective sizes required to achieve sufficient packets per spine to detect a desired drop rate with perfect precision. The last column indicates how many training iterations of Llama-3 70B are necessary to meet the required number of packets. 32, 64, 128 spines correspond to network sizes of 2k, 8k, 32k GPUs respectively. The data is extrapolated test bed measurements. [Testbed]}
    \label{tab:required-collective-sizes}
\end{table}

\begin{figure}
    \centering
    \includegraphics[width=.5\linewidth]{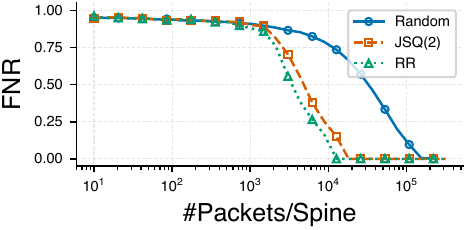}
    \caption{Random Packet Spraying increases the number of packets per spine necessary for accurate failure detection. RR refers to a per-priority level RR across all available spines. Despite the equal balance of RR, it only requires slightly fewer packets than JSQ(2). [Simulation]}
    \label{fig:random-lb}
\end{figure}

\paragraph{Random Packet Spraying}
The number of packets necessary for accurate detection depends on the load balancing strategy as shown in~\Cref{sec:design:threshold}. To showcase this behavior and to provide an estimate of the number of packets necessary with random packet spraying,~\cref{fig:random-lb} in \cref{appx:eval-robust} provides an analog to~\cref{fig:num-packets} comparing random spraying and JSQ(2).
Random packet spraying requires significantly more packets for failure detection due to the much higher variance compared to JSQ(2). Near perfect load balancing with a per-priority level RR results in an isolated RR of the measurement packets, requiring fewer packets than JSQ(2). \projname's time to detection increases with the variance of the packet distribution.

\subsection{Robustness} \label{sec:eval:robust}
We show that \projname is robust to non-pristine network environments, considering simultaneous gray failures, preexisting failures, competing traffic, network congestion, and packet reordering.

\paragraph{Multiple Gray Failures}
We observe no effect on the accuracy 
when injecting multiple simultaneous gray failures on up to 6\% of links on the path between two leaves (4 out of 64 links in a 32 spine topology) (\cref{appx:eval-robust}). However, for an unrealistically large fraction of affected links (e.g. 2 out of 16 links on the testbed), the accuracy is reduced.
The reason is that excessive retransmissions elevate the packet counters of some failed links above the detection threshold, causing false negatives.
However, \projname does \emph{not} produce false positives as no healthy link falls below the detection threshold.

\paragraph{Preexisting known failures}
Larger networks are likely to have preexisting disabled links which are reflected in routing tables. These cause network asymmetry, reducing the number of paths available for certain flows.
Affected flows must spray more packets over each remaining path.
As a result, increasing the number of preexisting failures \emph{improves} detection accuracy.
We observe in simulation that \projname maintains its accuracy under preexisting failures  (\cref{appx:eval-robust})

\paragraph{Network congestion}
Using a testbed experiment, we observe that \projname maintains its accuracy in face of congestion~(\cref{appx:eval-robust}).
Network congestion does not affect the spraying distribution as the distribution is controlled by the upstream path and the prioritization mechanism isolates measurements flows fully from competing traffic.
While the congestion control algorithm (CCA) may reduce the sending rate of a flow, this does not change the switch's spraying distribution if the flow is isolated for measurements by \projname. 
Since \projname aggregates packet counters over the lifetime of the flow, the detection is independent of the sending rate.
Since \projname accurately detects failures under congestion, it can operate in blocking networks.

\begin{figure}[t]
    \centering
    \subfigure[Jitter\simulation\label{fig:jitter}]{
    \includegraphics[width=.46\linewidth]{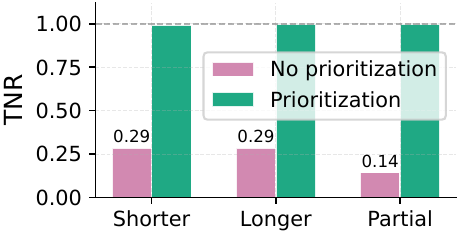}
    }
    \subfigure[Coverage\simulation\label{fig:coverage}]{
    \includegraphics[width=.46\linewidth]{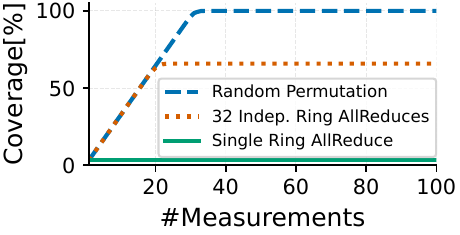}
    }
    \vspace{-1ex}
    \caption{(a) \projname requires prioritization in order to be unaffected by jitter. $1\%$ of links are assumed to be offline to create network asymmetry. (b)\projname selects flows to all available destination leaf switches, achieving the maximum coverage allowed by the workload.}
    \label{fig:robustness-coverage}
\end{figure}

\paragraph{Jitter of Competing Traffic}
As motivated in~\cref{fig:asymmetric-topology}, the timing of competing traffic in an asymmetric network impacts the packet spraying distribution.
However, as discussed in~\cref{sec:asymmetry-predictability}, prioritizing a flow during spraying restores the expected balanced distribution.
We confirm this behavior in simulation.
\Cref{fig:jitter} shows three scenarios of overlapping flows: First, a shorter flow both starts and completes while the measured flow is running. Second, the measured flow fully overlaps with a competing flow. Third, while the measured flow is running, a competing flow starts and continues until after the measured flow completes.
In all cases, without prioritization, the jitter between the two flows results in a low true negative rate (TNR), meaning that false positives are common.
When the measurement flow is prioritized, there are no false positives in any case.
We therefore conclude that prioritizing traffic during spraying is an effective mechanism to prevent competing traffic from interfering with \projname.

\paragraph{Packet Reordering}\label{eval:reordering} %
Network reordering may result in a situation where the measurement flow's last packet is not the final packet to traverse the destination leaf switch. Since the last packet causes the switch to stop counting, packets which arrive after the last packet will not be counted by \projname.
We experimentally confirm that under realistic conditions, the amount of reordering is too small to impact \projname's detection accuracy.
While the theoretical maximum reordering is approximately $7.5k$ packets\footnote{$\frac{\text{max. queuing delay}}{\text{packet pacing time}}$. With 3 queues each with 10 MiB max. queue size: $\lceil3\cdot\frac{10MiB}{25\frac{GB}{s}} \cdot (\frac{(4096 + 58)\frac{B}{Packet}}{25\frac{GB}{s}})^{-1}\rceil =7,573$ packets}, the amount of reordering in both simulations and the real world~\cite{loadbalancing-for-ai} is much smaller. In our worst-case simulation, two flows are sent from different source leaves to the same destination over 64 spines. To maximize reordering, we fully fail half of the uplinks such that each flow is sprayed via 31 non-overlapping spines, plus one single spine used by both flows. We observe a maximal reordering of 100 packets, far fewer than the $>100k$ total packets needed for failure detection in this network.

\subsection{Coverage}
\projname's flow selection policy aims to regularly check every path used by the current workload.
Because \projname uses application traffic to detect failures, it can only check paths used by the workload.
\Cref{fig:coverage} shows the percentage of destination leaves covered from a given source leaf as it selects flows for measurement.
For random permutation traffic, the source leaf rapidly covers all possible destinations, as flows are available to all of them.
For the 32 independent \ringallreduce workload, rings are independently and randomly selected, meaning that not all destination leaves have flows available for measurement. Still, \projname's flow selection policy covers all available destinations.
In the case of a single \ringallreduce, only one destination leaf switch is available. Nevertheless, because each leaf communicates with two different leaf switches, either as sender or receiver, failures can still be detected and localized.

\subsection{Impact on Application Performance}\label{sec:evaluation:assumptions}
\projname prioritizes \emph{a single flow} during spraying over all other spine-bound traffic in the source leaf. In a congested network, this may penalize competing flows.
To show that the effect on performance is negligible, we simulate a 32-spine topology in which 16 identically sized cross leaf flows are sent from the same leaf. To create network congestion, we disable two upstream links from the sending leaf, corresponding to a $3\%$ link failure rate in the entire cluster. Without prioritization, all flows are slowed equally. With a single prioritization, the prioritized flow speeds up by $0.2\%$ and all other flows slow down by $0.25\%$ relative to the non-prioritized scenario. 
The impact is low because each flow is sprayed across 30 paths, so the fraction of prioritized traffic on each path is at most $3.33\%$ of line rate, too small to have end-to-end impact on competing traffic. 

\subsection{Comparison with Path Probing}
\begin{table}[t]
\centering
\begin{tabular}{lrrrr}
\toprule
Drop rate & 8 spines & 16 spines & 32 spines & 64 spines \\
\midrule
0.5\% & 3,882k & 7,765k & 15,824k & 30,737k \\
1.0\% & 976k & 1,918k & 3,883k & 7,882k \\
1.5\% & 436k & 877k & 1,690k & 3,523k \\
\bottomrule
\end{tabular}
\caption{Probe packets required to detect a single-link failure (source-routed end-to-end probing, mean over trials, rounded to the nearest 1000) [Simulation]}
\label{tab:probe-packets-required}
\end{table}

Path probing is the state-of-the-art method for loss detection in DCNs~\cite{pingmesh, skeletonhunter, netbouncer}. The detection sensitivity depends on the number of probes sent, as more probes increase the probability of being dropped. The advantage of probing is that the system can choose the probe path using source routing.
\Cref{tab:probe-packets-required} shows the number of probing packets are required on average to detect an elevated drop rate on a single link. Large number of probes is necessary as the system needs test all paths individually and the testing must observe the probe being dropped on that path despite the low drop probability. 
We see that probing approaches introduce a significant additional overhead, whereas \projname achieves the same failure detection accuracy with only a single additional packet.

\section{Limitations and Future Work}

\paragraph{Weighted Packet Spraying}
Some AR spraying strategies do not distribute packets equally across all paths.
If packet-sprayed flows coexist with preexisting, non-packet-sprayed flows, even spraying may cause imbalanced load.
Weighted packet spraying allows certain queues to be preferred for spraying~\cite{weighted-packet-spraying}, to counter this effect.
If the weights are known by the receiving leaf, \projname's threshold calculation can use them.

\paragraph{Other Topologies}\projname focuses on 2-level topologies as they are increasingly used for large-scale training clusters following multiplane designs (\cref{sec:background}). In a 3-level Fat Tree, for example, \emph{two} spraying decisions are made: on the source leaf and on the upstream spine. \projname requires a \emph{single flow} to be prioritized at \emph{both} spraying decision. This means that all spine switches within a pod must prioritize the same flow. To satisfy this requirement with minimal coordination, one approach is to select a single leaf switch to choose a cross-pod flow for the entire pod to prioritize. One way to implement it is to pass a token between the leaves, such that only the bearer chooses cross-pod flows, ensuring that only one leaf select a cross-pod flow. This token is handed between the leaf switch in a round-robin order. As there is only a single prioritized cross-pod flow visible to the spine switches, all spine switches automatically isolate its spraying behavior. This approach guarantees that a single flow is isolated during all spraying decisions.
We leave the extension to other topologies for future work.

\paragraph{Detection of Access Link Faults}
Access links, which connect end hosts to leaf switches, are used by all paths of a flow.
As a result, \projname's approach, which measures what portion of a flow's traffic traverses each possible path, is not applicable to these links.
Instead, we sketch an extension to detect and localize gray failures on these links by counting the number of packets and \texttt{NACK}s observed during a flow.

If a packet is lost on the receiver's access link, it will have already passed through the destination leaf and counted by \projname. The retransmitted packet will also be counted, meaning that by the end of the flow, the sum of all per-spine counters is larger than the expected number of packets for the flow. This clear signal indicates a failure somewhere between counting the packet at the destination leaf and delivery at the destination NIC, usually on the access link.

If a packet is lost on the sending access link, it will not have been counted by the destination leaf yet. Only the retransmitted packets are counted,
meaning that both the distribution and sum of per-spine counters will be as expected, which excludes failures in the fabric or destination access link. However, the switch can still deduce the presence of a failure by counting the \texttt{NACK}s for a given flow. If the switch sees a large number of \texttt{NACKs} but excludes all other failures, it can conclude that the source access link must be failed.

\paragraph{Lossy Fabrics}
\projname detects gray failures in lossless networks through their resulting packet loss. In lossy fabrics, packet drops are normal and expected. \projname does not currently distinguish expected loss from gray failures, requiring significant adaptation to support lossy fabrics.

\section{Discussion}\label{sec:discussion}

\paragraph{Application use of network priorities}
Priorities are often used in networks. 
For example, MoE traffic may be prioritized over \dpallreduce due to its latency sensitivity~\cite{flowmoe, lina}.
While \projname reserves the highest priority for spraying isolation, network operators may still use the remaining priority levels.
Additionally, while a measured flow may be prioritized above another higher-priority traffic class, we show in~\cref{sec:evaluation:assumptions} that the performance impact is low.

\paragraph{Beyond \ringallreduce}
We evaluate \projname using \ringallreduce collectives since these are typically the largest collectives performed on scale-out network, 
making it easy for \projname to detect low-impact gray failures.
However, \projname can detect failures using \emph{any sprayed flow} as long as the length of the flow is known to the destination leaf switch.
In the case of small flows, data from multiple flows can be aggregated to achieve a strong signal.

\paragraph{Security}
We minimize \projname's attack surface by avoiding direct communications between application logic and control plane.
Switches parse flow announcement packets in the data plane, isolating the control plane.
In the worst case, an attacker can cause the control plane to perform the threshold computation on bogus values.
Further, the leaf switches do not change any packet contents beyond the measurement marker, which can be reset in the destination switch, ensuring that application traffic remains untouched.

\paragraph{Passive flow metadata extraction}
\projname's implementation requires a flow announcement packet informing the switches about the existing flows and their sizes. In principle, this information could be passively extracted from the existing traffic of the CCL, but we did not develop this capability.

\paragraph{System Integration}
\projname detects not only gray failures but all packet drops in the system. The network monitoring system (NMS) can cross-reference \projname's reports with existing tooling to prune alerts. After localizing a failure to a specific link, the NMS triggers a routing table update to mitigate the failure. A destination leaf switch alone is unable to localize the failure to a single link. However, it can still disable the entire faulty path locally. Since \projname requires on-switch deployment and full topology knowledge, it can only be used by the network operator.

\section{Related Work}\label{sec:related-work}

\textbf{Gray Failures} cause performance degradations with no individual system component ceasing to work entirely~\cite{fail-slow, bugs-in-the-cloud}. This makes them notoriously hard to detect~\cite{superbench}. Common root-causes are partial hardware failures or firmware bugs~\cite{bugs-in-the-cloud,singh2021surviving}. In networks specifically, packet buffer corruption~\cite{fail-slow}, high optical loss on fibers, or transceiver errors~\cite{zhuo2017understanding} can all cause packet loss which harms system performance but is not necessarily represented in switch counters~\cite{packet-level-telemetry}.

\noindent\paragraph{Failure Detectors}
\emph{Path Probing} approaches exhaustively check all network paths for reachability and low packet loss. PingMesh~\cite{pingmesh} is a popular choice for data center networks due to the low overhead of ICMP probes. NetBouncer~\cite{netbouncer} uses source routing to control the probing path.
R-PingMesh~\cite{r-pingmesh} extended PingMesh to test RDMA reachability, latency, and throughput. SkeletonHunter~\cite{skeletonhunter} optimizes the probing schedule to only validate paths that are used by the sparse communication matrix of the ML training jobs in the cluster.
\projname, in contrast, uses application traffic for measurements without requiring additional probe packets.

\emph{End-host detectors} validate either application metrics, such as collective completion time, or individual flow metrics~\cite{passive-failure-detection, 007}. To isolate individual link failures, network-path information must be available at the end hosts, making the approaches fundamentally incompatible with packet spraying.
SuperBench~\cite{superbench} tests end-to-end performance metrics to ensure the entire system performs within expectation.

\emph{In-network Telemetry (INT)} allows tracing of a packet's path through the network and collecting switch statistics in the data plane. This allows the comparison of counters and time information across switches, allowing the detection of abnormal delays or inconsistencies in packet counters. However, as gray failures are not necessarily represented in switch counters~\cite{packet-level-telemetry}, they cannot be relied upon for detection. INT further introduces end host involvement~\cite{pint, p4-int} and causes packet overheads, by either sending telemetry packets or introducing additional headers in application packets~\cite{p4-int}.

\emph{Control Plane Monitoring} systems collect packet counters, bandwidth usage, switch configurations, and protocol state from the switch control plane~\cite{snmp} and check for anomalies.
Control plane sampling relies on the correctness of switch counters, which cannot be taken for granted for gray failures~\cite{packet-level-telemetry}. 
The low time resolution of control plane sampling prevents correlating counter values across switches with high precision, preventing the detection of small failures.%

\textbf{Packet Spraying} is commonly used for multi path routing in ML training clusters since a per-packet load balancing decision results in more equalized path usage and thereby avoids link congestion. Switch-based packet spraying decides the upstream path on each network switch. DRILL~\cite{drill} uses a join-shortest-queue approximation~\cite{power-of-two-choices} to balance packets equally across all paths. Since randomly sampling queues and comparing their lengths is expensive at line rate, quantized adaptive routing~\cite{2006-adaptive-routing, dragonfly-plus} chooses the output queue randomly from all queues with a length under a given threshold.

Switch-based approaches remove the possibility to correlate a packet with a fixed network path, as was possible with ECMP. This allowed the application to change its path, e.g., to avoid congestion, by changing the flow-five tuple. REPS~\cite{reps} reintroduced this possibility using a set of host-controlled entropy vectors (EVs) to control the per-packet path. If a specific EV experiences congestion or packet loss, it is not reused. This approach allows packet spraying whilst retaining the possibility to exclude faulty paths. It avoids the oblivious spraying widely deployed in adaptive routing that \projname relies upon for failure detection.

\section{Conclusion}\label{sec:conclusions}
We present \projname, a gray failure detector for networks using adaptive routing. \projname works by passively analyzing the packet spraying distribution of application traffic and comparing it to an analytical prediction. For a network with 64 spines, \projname detects a link with a 1.5\% loss rate with perfect accuracy within a  single training iteration of the Llama-3 70B model, and a 0.5\% loss rate within 5 iterations. \projname does not introduce network overhead, covers all paths used by the application, operates entirely within switches with minimal application support, is robust to network noise caused by competing flows or congestion control, and can detect multiple parallel gray failures.

\section{Acknowledgments}
We thank our shepherd Nirav Atre and the reviewers for their helpful comments and feedback.
This work was supported by the Israel Science Foundation (980/21 and 1998/22) and by the Zuckerman STEM Leadership Program.

\bibliographystyle{plain}
\bibliography{main}

\appendix
\FloatBarrier
\crefalias{section}{appendix}

\section{Retransmissions in the prediction}\label{appx:retransmission}
\projname does not include retransmissions in the predicted spraying distribution as it increases the complexity of the detection while only slightly increasing the number of packets and thereby improving the measurement signal.

A retransmission of a RDMA write operation is indistinguishable from the first transmission. As a result the switches cannot automatically detect retransmission but need explicit markings by the sending host using additional notification packets. Avoiding double counting of packets if the packet was lost on the destination host link, would require explicit tracking of sequence numbers in the destination switch. 

Due to all these complexities and small effect on the detection result, we leave the inclusion of these effects for future work.

\section{Routing Algorithm}\label{appx:routing}
This section describes the assumed routing logic including the state required at each switch. \projname assumes a Fat Tree topology. Each switch has multiple routes to all destinations in its local routing table. We do not make assumptions about the routing protocol used to provision the tables. 
For a packet destined to a remote leaf, the leaf switch has routes to the destination via all spines. In AR, the source leaf chooses the route based on its local queue lengths. It will choose randomly among the possible routes with the shortest local queue length~\cite{spectrumx-arxiv}.

For calculating the expected number of packets on the destination switches, we assume that the routing tables are symmetric. If the destination switch has routes to the source switch via a certain set of spine switches, we assume that the source switch can use the reverse routes via same set of spines.

\section{Additional Experiments}\label{appx:eval-robust}
\begin{figure}
    \subfigure[Preexisting Failures\simulation]{
    \includegraphics[width=.46\linewidth]{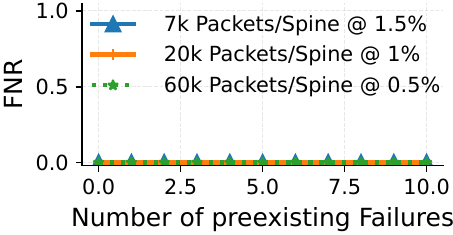}
    }
    \subfigure[Simultaneous Failures\simulation]{
    \includegraphics[width=.46\linewidth]{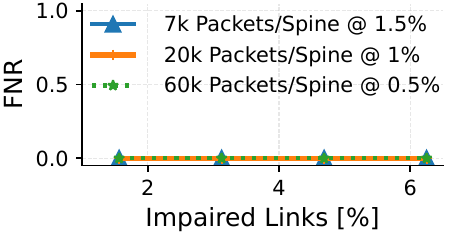}
    }\\
    \subfigure[Congestion Control\testbed]{
    \includegraphics[width=.46\linewidth]{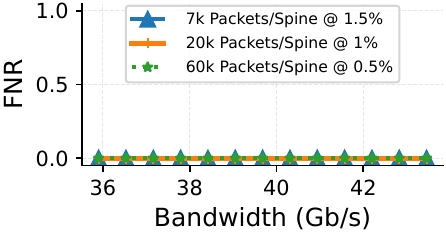}
    }\hfill
    \caption{Robustness results. The graphs show the claimed robustness to preexisting failures, simultaneous gray failures, and congestion control, claimed in~\cref{sec:eval:robust}}
    \label{fig:appx-robustness}
\end{figure}
\begin{table}
    \centering
    \small

    \begin{tabular}{c|c|cccc}
       \toprule
 \head{1cm}{Algo.} & \head{1cm}{$\frac{k Packets}{Spine}$}& Spines &  $k$Packets &\head{1.25cm}{Flow Size [GiB]} & Iter.\\
\midrule
\multirow{2}{1cm}{RR} & \multirow{2}{1cm}{$12$}& 32 & 406 & 1.55 & 0.20\\
&& 64 & 812 & 3.10 & 0.41\\
\midrule
\multirow{2}{1cm}{JSQ(2)} & \multirow{2}{1cm}{$18$}& 32 & 580 & 2.21 & 0.29\\
&& 64 & 1160 & 4.43 & 0.58\\
\midrule
\multirow{2}{1cm}{Random} & \multirow{2}{1cm}{$221$}& 32 & 7079 & 27.00 & 3.54\\
&& 64 & 14158 & 54.01 & 7.08\\
\bottomrule
    \end{tabular}
    \caption{Iterations required for different load balancing algorithms. Data is extrapolated from~\cref{fig:random-lb}.}\label{fig:num-packets}
    
\end{table}
\Cref{fig:appx-robustness} corroborates the robustness claims made in~\cref{sec:eval:robust}. This figure evaluates the effects of preexisting failures, simultaneous failures, and congestion control effects on false negative detection rates. For each case, we evaluate the ability to detect gray failures with 1.5\%, 1.0\%, and 0.5\% drop rates using 7k, 20k, and 60k packets per spine, respectively.

In all cases, the false negative rate is 0\%. Additionally, no false positives were observed. This confirms that \projname is robust to all three of these sources of measurement noise.

\section{Tofino Testbed Implementation}\label{appdx:testbed}
We implement the \projname prototype using the Intel Tofino 1 programmable switch platform for real world evaluations. Due to resource limitations, we emulate a 16-switch network using two physical switches: one to simulate 16 virtual switches, and one to inject failures.
The evaluation additionally uses a P$4_{16}$ implementation of a failure injector, adaptive routing, and congestion markings.

\paragraph{Switch Multiplexing}
We multiplex a single physical switch into 16 virtual vSwitches. Each virtual switch runs as its own instantiation of a control block.
Since our switch has fewer physical ports than our topology has virtual links between vSwitches, we multiplex physical links into virtual ones using VLANs.
When a packet arrives, it is dispatched to one of the vSwitches based on the packet's VLAN tag. The receiving vSwitch selects the next hop by changing the packet VLAN tag. Each virtual link between two vSwitches has one VLAN ID per direction. For example, when switch 1 wants to send a packet to switch 5, it sets the VLAN tag to $0x15$, while the tag $0x51$ is used for the reverse link.
The dispatcher invokes the vSwitch control block matching lowest 4 bits of the VLAN tag.

\paragraph{Shortest Queue Routing}
Implementing shortest queue packet spraying requires balancing across multiple output queues. However, since the testbed implements virtual switches, each vSwitch should ideally use its own independent set of queues for balancing. This is impossible to implement on the Tofino platform. Since our implementation only needs to mimic the spraying behavior and not the exact queuing delay, we use per-queue counters to approximate the spraying behavior of per-switch queues.
Due to switch multiplexing, packets traverse the same physical links, but are sprayed across different VLAN tags corresponding to different virtual upstream links.

We simulate the queuing in each vSwitch using per-VLAN packet counters. The counters are decremented at a constant rate by Tofino-generated timer packets.
Because of Tofino's memory access limitation, we split the logic for normal traffic into two, alternating on every packet: (1) We randomly sample two queue length counters and send the packet to the one with the lower counter. We store the chosen queue ID in a register. (2) We send the packet to the same queue ID as the previous and increment the counter for that queue.
This implementation drains the queues at a constant rate and increments them once for every two sent packets, while sending packets to the shortest out of two random queues.

\begin{figure}
    \centering
    \includegraphics[width=0.9\linewidth]{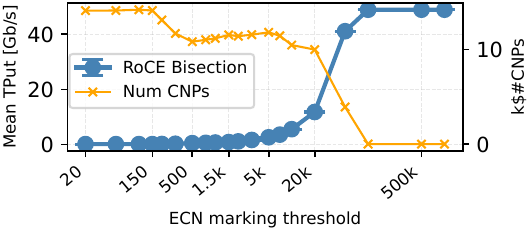}
    \caption{Throughput and number of congestion notification packets (CNPs) received during a RoCEv2 bisection bandwidth test with different ECN thresholds set on a virtual queue in the testbed. Increasing the threshold results in fewer CNPs and higher throughput.}
    \label{fig:testbed-linerate-control}
\end{figure}

\begin{table}[]
\centering
\begin{tabular}{c|ccc}
\toprule
Component & \projname  & \head{1.5cm}{SW Multiplexing} &  \head{1.5cm}{Basic SW Functions}\\
\midrule
Match Crossbar & 56 & 218 & 16 \\
SRAM & 63 & 156 & 10 \\
TCAM & 0 & 0 & 0 \\
VLIW Actions & 91 & 112 & 21 \\
Hash Bits & 180 & 2532 & 67 \\
Stateful ALUs & 27 & 0 & 0 \\
\bottomrule
    \end{tabular}
    \caption{Intel Tofino resource utilization.}
    \label{tab:tofino-resources}
\end{table}

\paragraph{Limiting Virtual Link Bandwidth}
Since all virtual cross-switch links share the same physical links, they can impact each other's bandwidth usage. For example, a high bandwidth between vSwitch 1 and 9 could congest a physical link shared with a flow between vSwitch 3 and 11, thus impacting that flow as well. To isolate the bandwidth usage of each virtual link to an equal share, we inject congestion markings on all packets on a virtual link to reduce the sending rate until the queue length drops below the marking threshold. \Cref{fig:testbed-linerate-control} shows the reduce sending rate of a RoCEv2 flow with congestion markings. Setting equal marking thresholds on all queue counters limits the sending rate of all flows independent of the path in the network.  

\paragraph{Failure Injection}
To simulate gray failures, we selectively drop a portion of packets with the VLAN tags corresponding to the link experiencing the gray failure.
We interpose each cross switch link in our testbed topology with another Intel Tofino switch running an adapted version of the failure injector used in~\cite{lumina}.
This system allows us to define a per-link packet drop rate, as well as to modify it at runtime.

\paragraph{Resource Consumption}
The data plane implementation consists of $1504$ LoC in P$4_{16}$ for both \projname and the entire testbed. The control plane implementation consists of $2611$ LoC in Python which set up routing tables, the entire \projname system, and a remote control interface to inject failures and read packet counter via the network for benchmarking. The data plane implementation requires 11 stages of the Tofino 1 pipeline. \Cref{tab:tofino-resources} details the resource consumption by component. Increasing the number of spines does not change the allocated SRAM due to the Tofino's block based allocation.

\end{document}